\documentclass[twocolumn,amsmath,amssymb,floatfix]{revtex4}

\usepackage{bm}
\usepackage{amsmath}
\usepackage{epsf}

\newcommand{\beq}{\begin{equation}}

\newcommand{\eeq}{\end{equation}}
\newcommand{\beqa}{\begin{eqnarray}}
\newcommand{\eeqa}{\end{eqnarray}}

\newcommand{\bn}{\hat{\bf n}} 
\newcommand{\bk}{{\bf k}} 
 
\newcommand{\bl}{{\bf l}} 
\newcommand{\dmm}{{\delta_m \delta_m}}
\newcommand{\dgm}{{\delta_g \delta_m}}
\newcommand{\dgg}{{\delta_g \delta_g}}
\newcommand{\de}{{\rm DE}}

\newcommand{\submitted}{{\rm submitted}}

\def\simlt{\lesssim}

\newcommand{\ApJS}{Astrophys. J Supp.}
\newcommand{\ApJL}{Astrophys. J Lett.}
\newcommand{\ApJ}{Astrophys. J}
\newcommand{\PRL}{Phys. Rev. Lett.}
\newcommand{\PRD}{Phys. Rev. D}
\newcommand{\MNRAS}{Mon. Not. R. Astron. Soc.}

\newcommand{\AsAs}{Astron. Astrophys.}
\newcommand{\AJ}{Astron. J.}
\newcommand{\amp}{\& }
\newcommand{\etal}{{\it et al. }}
\newcommand{\aut}[2]{{#2.\ #1,}}
\newcommand{\paut}[2]{{#2.\ #1} and}
\newcommand{\laut}[2]{{#2.\ #1,}}
\newcommand{\refs}[6]{#2, {\bf #3},  {#4} (#5).}

\newcommand{\mybib}[2]{\bibitem{#2}}

\begin{document}

\title{
Joint Galaxy-Lensing Observables and the Dark Energy}
\author{Wayne Hu$^{1}$ \& Bhuvnesh Jain$^{2}$}
\affiliation{
${}^1$Center for Cosmological Physics, Department of Astronomy and Astrophysics,
and Enrico Fermi Institute, University of Chicago, Chicago IL 60637 \\
${}^2$Department of Physics and Astronomy, University of Pennsylvania, Philadelphia,
PA 19104
}

\begin{abstract}
\baselineskip 11pt
Deep multi-color galaxy surveys with 
photometric redshifts will provide
a large number of two-point correlation observables:
galaxy-galaxy angular correlations, galaxy-shear cross correlations,
and shear-shear correlations between all redshifts.
These observables can potentially enable a joint determination of 
the dark energy dependent evolution of the dark matter and distances as
well as  the relationship between galaxies and dark matter halos.
With recent CMB determinations of the initial power spectrum, a measurement
of the mass clustering at even a {\it single} redshift will constrain a 
well-specified combination of dark energy parameters  in a flat universe;  
we provide convenient
fitting formulae for such studies.  The combination of galaxy-shear 
and galaxy-galaxy correlations can determine this amplitude at {\it multiple}
redshifts.  We illustrate this ability in a description of the galaxy 
clustering with 5 free functions of redshift which can be fitted from 
the data. The galaxy modeling is based on a mapping onto halos
of the same abundance that models a flux-limited selection.  
In this context and under a flat geometry, a 4000 deg$^{2}$ galaxy-lensing
survey can achieve a {\it statistical}
precision of $\sigma(\Omega_{\rm DE})=0.005$ for the dark energy
density,  $\sigma(w_{\rm DE})=0.02$ and $\sigma(w_a)=0.17$ for 
its equation of state and evolution, evaluated at dark energy
matter equality $z \approx 0.4$, as well as constraints
on the 5 halo functions out to $z=1$.  
More importantly, a joint analysis
can make dark energy constraints robust against 
systematic errors in the shear-shear correlation and halo modeling.

\end{abstract}
\maketitle

\section{Introduction}

In the successful standard cosmological model where structure in the universe
originates from Gaussian random density fluctuations in the initial conditions,
all statistical properties of cosmological structure  
observables depend on a single
quantity: the linear power spectrum of mass fluctuations.  The evolution of
this spectrum depends on the properties of the dark energy in a precisely
calculable way.   The task of extracting dark energy information 
from cosmological structures reduces to determining the relationship between
observables and the underlying linear mass power spectrum.

Deep, multi-color photometric galaxy surveys measure two sets of
cosmological observables: the angular distribution
of the galaxies and the weak lensing shear
induced on their shapes.
From these observables, three types of two-point
correlations can be constructed: the angular correlations
between the positions of the foreground galaxies, the shear-shear
correlations between background galaxies, and the galaxy-shear
cross correlations induced by the association of dark matter with
foreground galaxies.  

Whereas these correlations have so far been 
analyzed separately and/or with data from different surveys, 
the joint analysis of these measurements will be feasible from
forthcoming surveys. In this paper, we consider what can be learned from
the combined two point correlations. 
We shall see that
with the multitude of observables available, prospects for the 
joint determination of the cosmology and the relationship between galaxies
and mass are bright. 

Galaxy-shear cross correlations, also known as galaxy-galaxy lensing correlations,
were first detected by Brainerd \etal \cite{BraBlaSma96}, 
following earlier upper limits \cite{TysValJarMil84}.
The Red-Sequence Cluster and VIRMOS-DESCART surveys 
(\cite{Hoeetal02}, also \cite{SmiBerFisJar01}) are examples of the current
state of the art in deep surveys.  Though shallower,
the Sloan Digital Sky Survey (SDSS) complements these given
the larger population of foreground and background
galaxies \cite{Fisetal00,McKetal01,Sheetal03}.  These observations have been
interpreted in terms of the dark matter distribution associated with the
lensing galaxies and their environment under the halo model for galaxies
\cite{Sel00,GuzSel01,GuzSel02}.

Ongoing and future surveys such as the Canada France Hawaii Legacy 
survey, Pan-STARRS,
LSST, and SNAP \cite{Future} will produce 
multi-color catalogs of galaxies. 
Photometric redshifts can then be estimated
for these galaxies, allowing for measurements with foreground galaxies
extending to $z\sim 1$ and background galaxies up to twice as far in
multiple redshift bins. 

Photometric redshifts significantly augment the 
prospects for joint galaxy-lensing studies beyond the current state-of-the-art.   The number of observable cross-correlation
functions between the galaxies and the shear scales 
as the product of the number
of lens and source redshift bins and can easily exceed the number of
galaxy-galaxy clustering observables and independent
shear-shear correlation observables.

We consider prospects for constraining the evolution of the
dark energy through such surveys.  
The combination of galaxy-shear and galaxy-galaxy correlations is
particularly fruitful in that they allow for a joint solution of the
evolution of the matter and galaxy distributions.
These determinations can be cross checked against those from the shear-shear correlation.
The latter depend only on the mass power spectrum but are typically subject
to more severe systematic uncertainties.

We begin in \S \ref{sec:statistics} with a description of the statistical
methods employed.  These methods may be applied to any model of the
two-point correlations.  In the Appendices, we describe the parameterization
of the linear mass spectrum in the standard cosmological model, 
including convenient fitting functions for the dark energy effects,
and a generalization of the halo model for the association of galaxies
with the mass.
Our model for the latter 
is based on recent developments in galaxy simulations
which rely on matching the observed number density of galaxies
to the predicted number density of halos \cite{Kraetal03}.  
Utilizing these parameterized statistics,
we study prospects for constraining the properties of the dark energy in 
a spatially flat universe
in \S \ref{sec:darkenergy}.  We conclude in \S \ref{sec:discussion}.

\section{Statistical Methods}
\label{sec:statistics}

In this section, we describe the basic statistical approach to 
the joint study of galaxy and lensing power spectra.  These methods
are independent of the specific parameterization of the power spectra
described in the Appendix and employed in the following section.  
We begin with a brief review of
the relationship between angular and three dimensional power spectra
in \S \ref{sec:power}.  We relate these to the traditional galaxy-galaxy
lensing observables in \S \ref{sec:cross}.  In \S \ref{sec:fisher}
we describe how the sensitivity of power spectra 
to underlying parameters may be
quantified. 

\subsection{Power Spectra}
\label{sec:power}

Under the assumptions of statistical isotropy and small angles, 
the two-point observables
of a set of two dimensional scalar fields $x_i(\bn)$,
where $\bn$ represents the 
direction on the sky, is given by their angular power spectra
\begin{eqnarray}
\langle x_i^*(\bl) x_j(\bl') \rangle &=& (2\pi)^2 \delta(\bl-\bl') C_l^{x_i x_j}\,,
\end{eqnarray}
where ${\bl}$ is the Fourier wavevector or multipole
\begin{equation}
x_i(\bn) = \int {d^2 l \over (2\pi)^2} x_i(\bl) e^{i {\bl \cdot \bn}}\,.
\label{eqn:fouriertransform}
\end{equation} 
Here and throughout we use the same variable to represent the
field in angular and Fourier space.

Suppose these angular fields are related to three dimensional source fields
$s_i({\bf r};z)$ by a weighted projection
\begin{equation}
x_i(\bn) = \int dz W_i(z) s_i({\bf r}_i = \bn D_A; z) \,,
\end{equation}
where $D_A(z)$ is the angular diameter distance in comoving 
coordinates.  We will use semicolons to denote arguments that will
be suppressed where no confusion will arise. 

The Limber approximation \cite{Lim54,Kai92}
then relates the two dimensional power spectra
to the three dimensional power spectra as
\begin{equation}
C_l^{x_i x_j} = \int dz {H \over D_A^2} W_i(z) W_j(z) P^{s_i s_j}(k=l/D_A;z)\,,
\label{eqn:Limber}
\end{equation}
where $H(z)\equiv a^{-1} da/dt$ is the Hubble parameter and 
\begin{eqnarray}
\langle s_i^*(\bk) s_j(\bk') \rangle &=& (2\pi)^3 \delta(\bk-\bk') P^{s_i s_j}(k)\,,
\end{eqnarray}
defines the three dimensional source power spectrum.

For our purposes, the two dimensional fields will be the ``lens'' galaxy number density
fluctuations and the electric or $\epsilon$ component
of the weak lensing shear field measured with 
``source'' galaxies.  The lens and source
galaxies of a given survey may be divided into bins according to redshift, 
luminosity, color  or other criteria.  We will here concentrate on redshift binning.
However for notational convenience we will suppress the subscripts $i$ denoting
the bins for the rest of this section.
 
For the galaxy fluctuations, the source field is the three dimensional
number density $n_{V}({\bf r};z)$ or rather its fluctuations 
\begin{equation}
s({\bf r};z) = \delta_{g} = {\delta n_{V} \over \bar n_{V}}\,,
\end{equation} 
and the weight for the angular fluctuation field
$g(\bn)$ is the normalized redshift distribution function 
\begin{equation}
W_g(z) = {D_A^2 \over H} {\bar n_V \over \bar n_A} \,,
\label{eqn:wg}
\end{equation} 
where the normalization factor 
\begin{eqnarray}
\bar n_{A} &=& \int dz {D_A^2 \over H} \bar n_V
\label{eqn:na}
\end{eqnarray}
is the angular number density in ${\rm sr}^{-1}$.  Note that
the weights are normalized so that $\int W_g(z) dz= 1$.  

For the weak lensing shear, the underlying scalar field is the electric 
component of the shear field $x(\bn) = \epsilon(\bn)$
which manifests itself as a shearing of background galaxy
images according to the complex shear
\begin{equation}
\gamma_1(\bn) \pm i \gamma_2(\bn) = \int {d^2 l \over (2\pi)^2} 
\epsilon(\bl) e^{\pm 2i \phi_l} e^{i \bl \cdot \bn}\,,
\label{eqn:shear}
\end{equation}
where $\phi_{l}$ is the azimuthal angle of the Fourier vector with respect
to the $\hat{\bf e}_{1}$ axis.
The $\epsilon$ field itself
is a projection of the mass density fluctuation 
\begin{equation}
s({\bf r};z) =\delta_m = {\delta \rho_m \over \rho_m}\,,
\end{equation}
and hence is equal to the convergence $\kappa(\bn)$.

The weights are given by the efficiency for lensing a 
population of source galaxies 
\begin{equation}
W_\epsilon(z) = {3 \over 2}\Omega_m {H_0 \over H} {H_0 D_{OL} \over a}
\int_z^\infty dz' {D_{LS}\over D_{OS}} W_g(z')\,,
\label{eqn:efficiency}
\end{equation}
where the angular diameter distance to the lens is $D_{OL}=D_A(D)$,
to the source $D_{OS}=D_A(D')$ and between the lens and the source
$D_{LS}=D_A(D'-D)$.  Here $D(z)$ is the comoving coordinate distance and
note that $D=D_A$ in a flat universe.
The distribution of source galaxies $W_g(z')$ need not
be the same as for the (lens) galaxies above.  
Furthermore $W_g$, the normalized redshift distribution, is the direct observable so that the efficiency
$W_\epsilon$ for a known $W_g$ may be used to probe cosmology.  

The complete two point statistics of the shear and galaxy correlations
are thus specified by a choice of cosmology and a description of the
underlying three dimensional power spectra 
$P^{\delta_m \delta_m}$, $P^{\delta_g \delta_g}$ and
$P^{\delta_g \delta_m}$ as a function of wavenumber $k$ and redshift $z$.
The latter two will depend not only on cosmology but also on
galaxy properties.

\subsection{Cross Correlation Functions}
\label{sec:cross}

The galaxy-shear cross power spectra are related to the more
familiar cross correlation functions through the Fourier transform
relations Eqn.~(\ref{eqn:fouriertransform}), (\ref{eqn:shear})
\begin{eqnarray}
\langle \gamma_1(\bn) \delta_g(\bn') \rangle 
&=& \int {d^2 l \over (2\pi)^2} C_l^{g \epsilon} \cos (2\phi_l)
	e^{i \bl \cdot (\bn-\bn')} \,,\nonumber\\
&=& -\int {ldl\over 2\pi} C_l^{g\epsilon} \cos(2\phi) J_2(l\theta)\,,
\nonumber\\
\langle \gamma_2(\bn) \delta_g(\bn') \rangle 
&=& -\int {ldl \over 2\pi} C_l^{g\epsilon} \sin(2\phi) J_2(l\theta)\,.
\label{eqn:pixelcorrelation}
\end{eqnarray}
where $\bn-\bn'=(\theta,\phi)$ is the 
angular separation vector with magnitude $\theta$ and azimuthal angle
$\phi$ with respect to the $\hat {\bf e}_1$ axis.
We have used the identity
\begin{equation}
e^{i\bl \cdot (\bn-\bn')} = J_l(l\theta) + 2\sum_{m=1}^{\infty} i^m J_m(l\theta)
\cos[m(\phi_l-\phi)]\,.
\end{equation}
Note that the correlation functions depend 
on both the magnitude
of the separation vector $\theta$ and the azimuthal angle $\phi$.
Despite this complication, Eqn.~(\ref{eqn:pixelcorrelation}) can be straightforwardly
used to generalize a maximum likelihood
estimator for the shear-shear angular power spectrum (e.g. \cite{HuWhi00,Pen03}).

Although the galaxy-shear cross power spectrum is thus a direct observable,
observations to date have focused on the tangential shear
component around galaxies due mainly to systematic effects in
the shear measurement.  It is therefore useful to relate the two approaches.

We can express the tangential shear about a galaxy at the origin as
\begin{equation}
\gamma_T(\bn) = -\gamma_1(\bn) \cos(2\phi) - \gamma_2(\bn) \sin (2\phi)\,.
\end{equation}
The angular correlation function then becomes a function of $\theta$ alone 
and is given by
\begin{eqnarray}
\langle \gamma_T(\theta) \rangle_{\rm halo} &\equiv&
\langle \gamma_T(\theta)\delta_g(0) \rangle \nonumber\\
&=& \int {ld l \over 2\pi} C_l^{g\epsilon} J_2(l\theta)\,.
\label{eqn:tangential}
\end{eqnarray}
Under the ergodic assumption, this quantity can be reinterpreted as
the average tangential shear around lens galaxies in the sample volume.

For a narrow redshift distribution of source and lens galaxies, the
tangential shear directly probes the galaxy-mass power spectrum at the
lens redshift.  Substitution of the Limber equation (\ref{eqn:Limber}) in
Eqn.~(\ref{eqn:tangential}) gives 
\begin{align}
\langle \gamma_T({R\over D_{OL}}) \rangle_{\rm halo}
&\equiv {\Delta \Sigma(R) \over \Sigma_{\rm cr}  } \\
&= {\rho_0 \pi \over H_0 \Sigma_{\rm cr}} \int {dk \over k} \left( {H_0 \over k} \right) 
{k^3 P^{\delta_g\delta_m} \over 2\pi^2}
J_2(k R) \,, \nonumber
\end{align}
where $\rho_0 / H_0 = 832 \Omega_m h M_\odot$ pc$^{-2}$ and $R$ is
the distance transverse to the line of sight. 
The critical surface density is given by
\begin{eqnarray}
\Sigma_{\rm cr}^{-1} & = & {4 \pi G \over c^2} {D_{OL} D_{LS} \over a_L D_{OS} }=
{3 \over 2} {H_0^2 \over \rho_0} \Omega_m { D_{OL} D_{LS} \over a_L D_{OS} }\,.
\end{eqnarray}
For a single lens galaxy with an azimuthally symmetric density profile $\rho(R,D)$, these quantities
are related to the projected or surface mass density
\begin{equation}
\Sigma(R) = \int d D \rho(R,D)\,,
\end{equation}
through
$\Delta \Sigma(R) = \bar \Sigma(R) - \Sigma(R)$, where
 the average is over transverse distances interior to $R$.
Note that all distances are in comoving coordinates.

The tangential shear technique throws away information by combining
the two components of the shear before averaging. 
It contains the majority of the
signal when considering a spherically symmetric density 
distribution about a galaxy.
At large radii, the usual approach is
to attempt a reconstruction of the scalar $\epsilon(\bn)$ 
($\equiv \kappa(\bn)$, the convergence in the weak lensing limit) 
out of
both components (e.g. \cite{KaiSqu93}).  This reconstructed field then acts 
as a template density map for the correlation
\begin{eqnarray}
\langle \epsilon(\theta) \delta_g(0) \rangle &=&
\int{d^2 l \over (2\pi)^2} C_l^{g \epsilon} e^{i \bl \cdot \bn} \nonumber\\
&=& \int {ld l \over 2\pi} C_l^{g\epsilon} J_0(l\theta)\,.
\end{eqnarray}
Alternately, one can go the other way and define a template shear
\cite{BerJai03}
\begin{eqnarray}
\gamma_1^{g}(\bn) \pm i \gamma_{2}^{g}(\bn) &=& \int {d^2 l \over (2\pi)^2} \delta_g(\bl) e^{\pm  2i\phi_l} 
	e^{i \bl \cdot \bn} \,,
\end{eqnarray}
and construct the correlation
\begin{eqnarray}
\langle 
\gamma_1^g(\theta) \gamma_1(0)
+ \gamma_2^g(\theta) \gamma_2(0)
\rangle
&=& \int {ldl \over 2\pi} C_l^{g\epsilon} J_0(l\theta)\nonumber\\
&=& \langle \epsilon(\theta) \delta_g(0) \rangle \,.
\label{eqn:sheartemplate}
\end{eqnarray}
Formally the two techniques construct the same correlation 
function and preference for one versus the other is a matter
of considering errors in the reconstruction.

\subsection{Fisher Matrix}
\label{sec:fisher}

Given angular power spectra that are defined by a set of cosmological
and galaxy parameters $p_\alpha$, 
forecasts on how well such parameters can be extracted from the data
is an exercise in error propagation.

In general, the observed two points statistics of the
angular fields $\tilde C_l$ will receive
a contribution from noise sources which we will assume to be statistically
isotropic
\begin{equation}
\tilde C_l^{x_i x_j} = C_l^{x_i x_j} + N_l^{x_i x_j}\,.
\end{equation}
We will further assume that the noise contributions for the galaxy and shear
fields arise from uncorrelated shape and shot noise
\begin{eqnarray}
N^{\epsilon_i \epsilon_j}_l &=& \delta_{ij} {\gamma_{\rm rms}^2 \over \bar n_{A i}}\,,
\nonumber\\
N^{g_a g_b}_l &=& \delta_{ab} {1 \over \bar n_{A a}}\,,\nonumber\\
N^{\epsilon_i g_a}_l &=& 0\,,
\label{eqn:noise}
\end{eqnarray}
where $\gamma_{\rm rms}$ is the rms shear in each component arising from
the intrinsic ellipticity of the galaxies and measurement noise.

The shear fields are expected to be nearly Gaussian with respect
to power spectrum errors for $l \simlt 10^3$ due to linearity and 
projection \cite{ScoZalHui99} as borne out in simulations \cite{WhiHu00}.  
For the galaxy fields, the transition scale
is at somewhat lower $l$ depending on the width of the projection 
(e.g. \cite{EisZal01,ScoShe01}).
However for both, the noise contributions at $l \gg 10^3$ dominate the errors
and mask the non-Gaussianity of the underlying fields.
Under the assumption of Gaussian statistics for the fields, 
the information contained in the power spectra can be quantified by the
Fisher matrix 
\begin{equation}
F_{\alpha \beta} =  f_{\rm sky} \sum_l { (2l+1)\Delta l \over 2}
\,{\rm Tr}[ 
{\bf D}_{l\alpha}  \tilde{\bf C}_{l}^{-1}
{\bf D}_{l\beta}  \tilde{\bf C}_{l}^{-1}
]\,,
\label{eqn:Fisher}
\end{equation}
where the sum is over bands of width $\Delta l$ in the power spectrum and
$f_{\rm sky}$ is the amount of sky covered by the survey.  
The rough $f_{\rm sky}$ scaling is valid for contiguous regions with comparable extent
in each of the two angular directions.

Here 
we have suppressed the $(i,j)$ indices in a matrix notation and
\begin{equation}
[{\bf D}_{l\alpha}]^{ij} \equiv
D_{l\alpha}^{x_i x_j} = {\partial C_l^{x_i x_j} \over \partial p_\alpha}\,.
\label{eqn:derivs}
\end{equation}
The inverse Fisher matrix approximates the covariance matrix of the
parameters ${\bf C}^p \approx ({\bf F}^{-1})$.

One can also break this information down into subsets
\begin{equation}
F_{\alpha \beta}^{\rm sub} =  
f_{\rm sky} 
\sum_l { (2l+1)\Delta l }
\sum_{(ij)(mn)}
{D}^{x_i x_j}_{l \alpha} [\tilde{\bf C}_l^{\rm sub}]^{-1} {D}^{x_m x_n}_{l \beta}
\label{eqn:Fishersub}
\end{equation}
where sum over $ij$ and $mn$ pairs can run over a restricted subset of
the observables. The covariance matrix of the subsetted power spectra is
given by 
\begin{equation}
[\tilde {\bf C}^{\rm sub}_l]^{ij,mn}= \tilde C_l^{x_i x_m} \tilde C_l^{x_j x_n} +
\tilde C_l^{x_i x_n} \tilde C_l^{x_j x_m}. 
\label{eqn:crosscov}
\end{equation}
In the limit that the sum is over all combinations, Eqn.~(\ref{eqn:Fishersub})
returns Eqn.~(\ref{eqn:Fisher}).   This subsetting also clarifies
the role of the Gaussian assumption.  Gaussianity implies a diagonal
covariance matrix in $l$ and reduces its form to the product of power spectra
in Eqn.~(\ref{eqn:crosscov}). 
Non-Gaussianity from non-linear structure formation correlates the power
between high $l$ bands  but in a fairly simple way: all bands share
a common normalization whose variance is determined not by Gaussian
statistics but by the sample variance near the non-linear scale.  
Under the halo model of Appendix B, 
this behavior arises because the shape reflects the shape of halo profiles
whereas the amplitude reflects their abundance.  This abundance fluctuates
with the local mean density. 

Under the Limber approximation of Eqn.~(\ref{eqn:Limber}),  given $N_L$ 
 lens galaxy samples in disjoint
redshift bins and $N_S$ source galaxies samples, there are $N_L$ distinct galaxy spectra,
$N_S (N_S+1)/2$ shear spectra, and $N_L N_S$ galaxy-shear cross spectra.
The potentially large number of cross spectra offers great opportunities
for studies of galaxy evolution and cosmology.
Consider then the Fisher matrix of cross spectra alone
\begin{equation}
F_{\alpha \beta}^{g\epsilon} =  f_{\rm sky}
\sum_l { (2l+1)\Delta l }
\sum_{(ai)(bj)}^{N_L N_S}
{ \partial C_l^{g_a \epsilon_i} \over \partial p_\alpha }
[\tilde {\bf C}_{l}^{g\epsilon}]^{-1} 
{ \partial C_l^{g_b \epsilon_j} \over \partial p_\alpha }
\label{eqn:Fishercross}
\end{equation}
where
\begin{equation}
[\tilde{\bf C}_l^{g\epsilon}]^{ai,bj}= 
\tilde C_l^{g_a g_b} \tilde C_l^{\epsilon_i \epsilon_j} +
\tilde C_l^{g_a \epsilon_j} \tilde C_l^{g_b \epsilon_i}\,.
\end{equation}
Note that the total variance 
of the shear and galaxy fields contributes to the noise of the cross 
correlation as a type of sample variance.  
Furthermore in the low signal-to-noise regime, the covariance
is dominated by the product of power spectra of the galaxy and shear fields not the
sample variance of the signal.   In this case, the $l$-diagonal form 
of the Fisher matrix
depends on the assumption of statistical isotropy and 
remains valid even when the shear and galaxy fields are strongly 
non-Gaussian.

The framework described here is general and may be applied to any parameterized model for
the underlying three dimensional power spectra $P^{\dmm}$, $P^{\dgm}$ and $P^{\dgg}$ 
and any selection criteria for the galaxies.   In Appendix A 
we describe
the well-tested standard model for the linear mass spectrum and how it
depends on the dark energy.  In Appendix B 
we develop the halo model for the galaxy and cross spectra.  Motivated by recent simulations
which
associate galaxies with substructure in dark matter halos, we utilize 5 free functions
of redshift to describe the occupation of galaxies in dark matter halos.  
By discretizing these functions into observed redshift bins we obtain a halo model
parameterization with $5 N_L$ parameters

Because even this multi-dimensional halo
model may not be sufficiently realistic, we will attempt in the next section to
separate cosmological information that does and does not depend on the details of the
 halo model.
For instance in the large scale regime where galaxies clustering is nearly fully correlated
with the mass, a measurement of the galaxy auto and cross power spectra is
essentially a measurement of the mass power spectrum.  Furthermore, the
angular cross spectra as a function of source galaxy redshift scale with distance in 
a known way through the lensing efficiency for any choice of the underlying three
dimension power spectrum $P^{\dgm}$.

\section{Dark Energy Prospects}
\label{sec:darkenergy}

In this section, we study the prospects for dark energy constraints with
galaxy-lensing power spectra.  We begin in \S \ref{sec:fiducial} by defining
the fiducial survey and the galaxy selection.  In \S \ref{sec:JT} we
study the distance related or halo model independent information in the
galaxy-shear cross correlation and in \S \ref{sec:full} the joint constraints
from all power spectra.

\subsection{Fiducial Survey}
\label{sec:fiducial}

For illustrative purposes, let us define a fiducial survey that is
loosely based on 
a next generation ground based lensing survey.   We take a source redshift
distribution of the form
\begin{eqnarray}
W_{g;S} \propto z^2 e^{-(z/z_{W})^2}
\end{eqnarray}
with $z_{W}$ corresponding to a median redshift $z_{{\rm med}; S}=1.5$ 
and an angular number density of 
$\bar n_{A}=70$ arcmin$^{-2}$.    This corresponds roughly to a magnitude limit
of $I \approx 27$.  For the shape noise 
we take a shear rms per component
of $\gamma_{\rm rms}=0.3$ which reflects the intrinsic ellipticity and
measurement errors of ground based observations (see e.g. \cite{SonKno03}). 
Note that the noise
variance scales as $\gamma_{\rm rms}^{2}/\bar n_{A}$ in 
Eqn.~(\ref{eqn:noise}).
We take a survey area of 4000 deg$^{2}$.

From the survey galaxies we choose a galaxy lens population from the
high luminosity tail.   To balance signal strength and halo model
robustness per lens galaxy against
lens abundance, we choose lens galaxies with an abundance 
corresponding to a mass threshold of 
$M_{\rm th}=10^{13.5} h^{-1} M_\odot$ in the fiducial model.
Due to this tradeoff, the net signal-to-noise ratio is only weakly dependent
on the threshold.   Note that the mass threshold is not held fixed as
halo and cosmological parameters are varied (see Appendix B) 
The fiducial mass threshold
simply defines the redshift distribution and angular density of the
lens galaxies.  These are the quantities held fixed under the variations.

For the redshift binning, we typically choose
between 2-5 photometric redshift bins
in the source distribution.  Further divisions do not substantially
enhance constraints on 
the dark energy given the broad efficiency function \cite{Hu99b}.
These are taken to have equal extent in 
redshift out to $2 z_{{\rm med};S}$ but with the last bin containing the remaining high
redshift galaxies.  For the lens galaxies,
we limit the populations to $z < 1$ since they will require more accurate
photometric redshifts.
We typically
take $N_L=10$ lens galaxy bins reflecting a photometric redshift 
accuracy of $\Delta z=0.1$.  The total lens population then has an
 angular number density of $0.026$ gal arcmin$^{-2}$ with
a median redshift of $z_{{\rm med};L}=0.7$. 

Finally, we allow for uncertainties in the
shape of the mass power spectrum and initial normalization
by taking priors of $\sigma(\ln \delta_\zeta)=\sigma({n_s})
=\sigma(\ln \Omega_b h^2)=\sigma(\ln \Omega_m h^2)=0.1$ 
corresponding to a conservative interpretation of
current constraints (see Appendix A and e.g. \cite{Speetal03}).
With the fiducial lens survey, dark energy results depend only
weakly on these prior assumptions.
Note that we take no prior constraints on the dark energy parameters so that
the projected constraints reflect only the potential of galaxy-lensing
power spectra.

\begin{figure}[tb]
\centerline{\epsfxsize=3.0in\epsffile{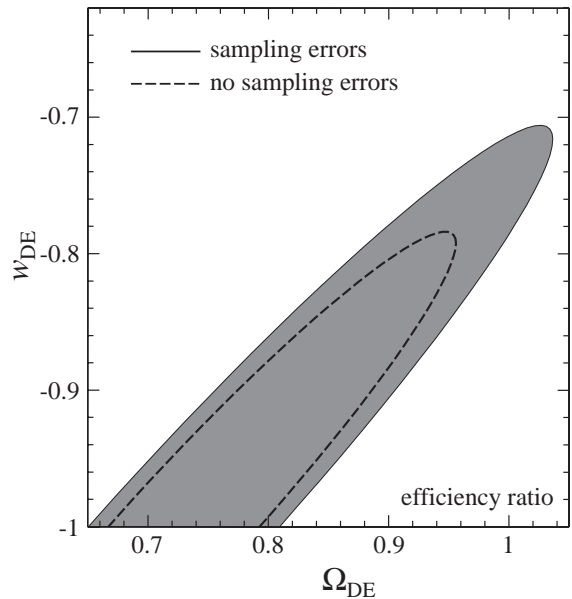}}
\caption{\footnotesize Lensing efficiency ratio and 68\% CL forecasts
on a two parameter
model for the dark energy ($\Omega_{\rm DE}$, $w_{\rm DE}$).   
With two or more source galaxy populations,
here $N_S=2$, the ratios of galaxy-shear power spectra provides
a measure of distances through the efficiency ratio.  Here
we take $N_L=10$ lens galaxy populations and marginalize the
amplitude of the galaxy-mass power spectrum in 
all bandpower and redshift bins.   The inner contour 
shows the effect of neglecting the sampling errors and shows that
sample and shot noise are comparable for a shear noise of 
$\gamma_{\rm rms}=0.3$ and a source density of $\bar n_{A}=70$
arcmin$^{-2}$. These parameters and a  4000 deg$^2$ survey are 
assumed here and throughout.} 
\label{fig:sample}
\end{figure}

\subsection{Model Independent Constraints}
\label{sec:JT}

Comparing lensing observables for different 
source redshifts has been proposed as a way of measuring distances with galaxy clusters
\cite{LinPie98,GauForMel00,GolKneSou02,Ser02}.   Recently, wide field weak lensing statistics
have also been developed to exploit this test \cite{JaiTay03,BerJai03}.
In the limit of
infinitesimal lens galaxy redshift
bins, i.e. in
Eqn.~(\ref{eqn:efficiency}) 
\begin{equation}
W_{g;L} \rightarrow \delta(z-z_L)\,,
\end{equation}
the ratio of the galaxy-shear
cross correlation in multiple shear source bins depends
only on the ratio of efficiencies. This
in turn depends only on angular diameter distances and redshifts.  
In terms of the cross power spectra
\begin{equation}
{ C_l^{g \epsilon_1} \over C_l^{g \epsilon_2}} 
= {W_{\epsilon_1}(z_L) \over W_{\epsilon_2}(z_L)} \,.
\end{equation}
Hence the galaxy-shear cross correlation
provides information on  the dark energy which
is immune to uncertainties in the modeling of the underlying galaxy-mass correlation.
The ratio is also immune to sample variance in the {\it signal}, i.e. galaxy
to galaxy variations in the underlying mass correlation 
and hence the non-Gaussianity
of the signal.  In the Fisher matrix of Eqn.~(\ref{eqn:Fishercross}), if the noise
terms $C_l^{gg}$ and $C_l^{\epsilon\epsilon}$ vanish, the covariance matrix becomes
singular implying infinitesimal errors in distance parameters.  In fact the 
technique works for individual galaxies in principle.  Unfortunately with
realistic noise estimates, the tradeoff between model independence and 
sensitivity is severe.  For example, a 10\% change in the equation of
state parameter $w_{\rm DE}$ typically yields an $\sim 0.1\%$ change in the 
efficiency ratio 
and so measurement of the effect is only possible with 
large galaxy and shear surveys.
Contrast this with the several percent change in the absolute growth or
shear amplitude implied by Eqn.~(\ref{eqn:growthfunction}) and
illustrated in Fig.~\ref{fig:param}.

Even for this statistic the halo model enters in three ways: 
by defining the strength of the signal, the
sample variance of the {\it noise} and the
accuracy to which the redshifts of the galaxy lenses must be known.  
The sample variance arises from contributions to the shear 
from structure along the line of sight not
associated with the galaxy population and from the intrinsic clustering of
galaxies.   
Furthermore, with
finite-width redshift bins in the lens galaxy distributions, the 
efficiency ratios depend on the model for the evolution of the underlying power
spectra across the bins \cite{BerJai03}.   
Even with our 5 halo parameter model for the evolution in each bin, 
the constraints are compromised.  No constraints are possible with
photometric redshift bins of $\Delta z \approx 0.1$ if
we marginalize all 5 parameters for a given bandpower in $l$.
By combining multiple bandpowers, one can recover dark energy information,
but that amounts to utilizing information from the shape of the underlying 
correlation and again degrades the model independence of the effect.
We will return to this type of constraint in the next section.

To study the efficiency ratio test, 
we instead marginalize a {\it single} halo parameter per redshift and bandpower
bin.  This model is equivalent to marginalizing a
constant amplitude or bias per bandpower in the underlying power spectrum.  
Hence it eliminates information from the shape but retains an assumption on
the evolution of the galaxy-mass power spectrum.
Results do not
depend on the number of bands employed so long as the signal and noise 
dominated regimes are separated 
since the additional parameters are employed simply
to remove shape information from measurements of different bands.
We take the parameter to be the
satellite normalization $A_s$ for convenience and have
verified that the results do not depend on this choice. 

In Fig.~\ref{fig:sample} we show the constraints in the constant equation of
state, dark energy density  
($w_{\rm DE}$-$\Omega_{\rm DE}$)
plane for $N_L=10$, $N_S=2$ (see Appendix A 
for dark energy
parameter definitions).  Note that when moving to a varying $w$, the errors
on equation of state at the best constrained redshift $w_{\rm pivot}$ remain 
the same $\sigma(w_{\rm pivot}) = \sigma(w_{\rm DE})$,
whereas those on $\Omega_{\rm DE}$ increase.
We can also assess the effect of sample variance by artificially removing
the terms in the covariance (\ref{eqn:crosscov}) that are proportional
to $C_l^{\epsilon\epsilon} C_l^{gg}$.  This causes an improvement in the
errors by $\sim \sqrt{2}$ indicating that the sample variance is comparable to
shot variance.  This near equality reflects the fact that these halos 
have a projected scale
radius of order $1'$ (see Fig.~\ref{fig:gamt}).   
On these scales, the assumed shape noise 
$\gamma_{\rm rms}/ \sqrt{ \bar n_{Ag}}\sim 0.02$ is comparable to the cosmic shear.
We include the information from all scales; in practice the signal to
noise has converged well before our numerical maximum $l=30000$.

The discrepancy of our results with those of \cite{JaiTay03} is due to 
a factor of two error in the calculation of the signal (since corrected, astro-ph/0306046 v3), 
to a more realistic model for halo profiles, and the inclusion of
sampling errors.  Comparison of the results 
also requires implementing their prior on $\Omega_{\rm DE}$ and
their shape noise specifications. 
These results agree with an independent
and concurrent study which also extended the model-independent techniques
to shear-shear correlations \cite{ZhaHuiSte03}. 

Raising the number of source ($N_S$) and lens ($N_L$) divisions or choosing less rare galaxy tracers
only slightly improves these constraints.  Note that the latter
entails employing less massive objects with higher number density but 
weaker shear signal.
Finally note that the template technique of \cite{BerJai03} can be reexpressed
as a measurement of the zero lag correlation of Eqn.~(\ref{eqn:sheartemplate})
or a single bandpower
\begin{equation}
\gamma_{\rm rms}^{g} = \int{d^2 l \over (2\pi)^2} C_l^{g\epsilon}\ .
\end{equation}

Although the efficiency ratio constraints are relatively weak,
they are fairly robust to both the halo model and non-Gaussianity
of the correlation on small scales.   Moreover, they 
become more powerful when combined
with complementary probes of shape and amplitude of the various power
spectra on large scales as we shall now see.

\begin{figure}[tb]
\centerline{\epsfxsize=3.0in\epsffile{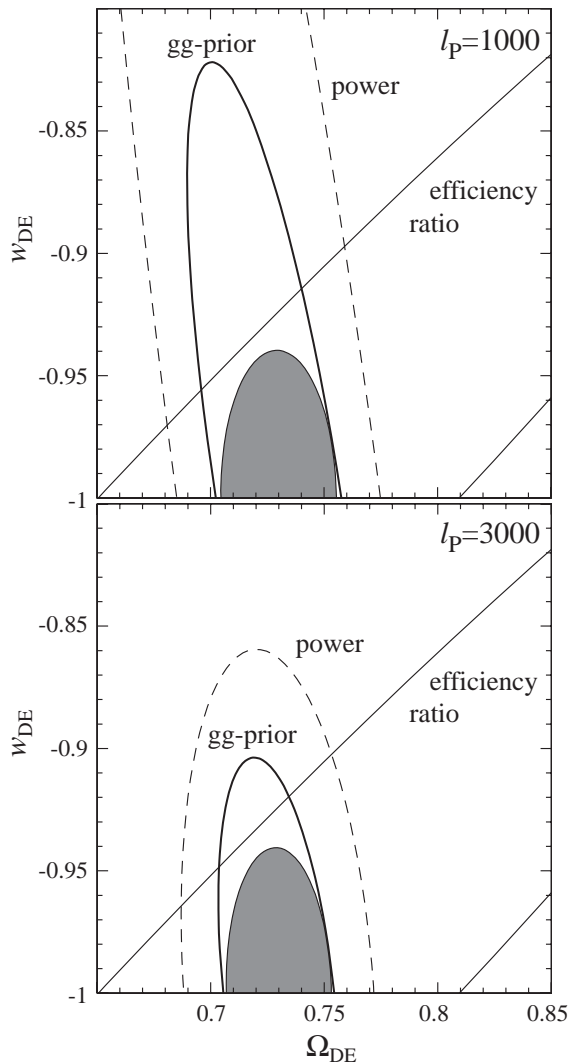}}
\caption{\footnotesize
Galaxy-shear power spectrum 68\% CL constraints on a two parameter dark
energy model ($\Omega_{\rm DE}$, $w_{\rm DE}$) with power spectrum
information out to $l_P=1000$ (upper) and $3000$ (lower). 
Galaxy-shear constraints (dashed lines) are complementary to the
efficiency ratio test (this solid) and are assisted by the addition of
galaxy-galaxy constraints (thick solid) which help determine the $5 N_L=50$
halo parameters.
Joint constraint (shaded) is only 
weakly dependent on $l_{P}$ and hence non-Gaussian 
errors. Here the number of source redshift distributions $N_S=2$.
}
\label{fig:cross}
\end{figure}

\subsection{Model Dependent Constraints}
\label{sec:full}

\begin{figure}[tb]
\centerline{\epsfxsize=3.0in\epsffile{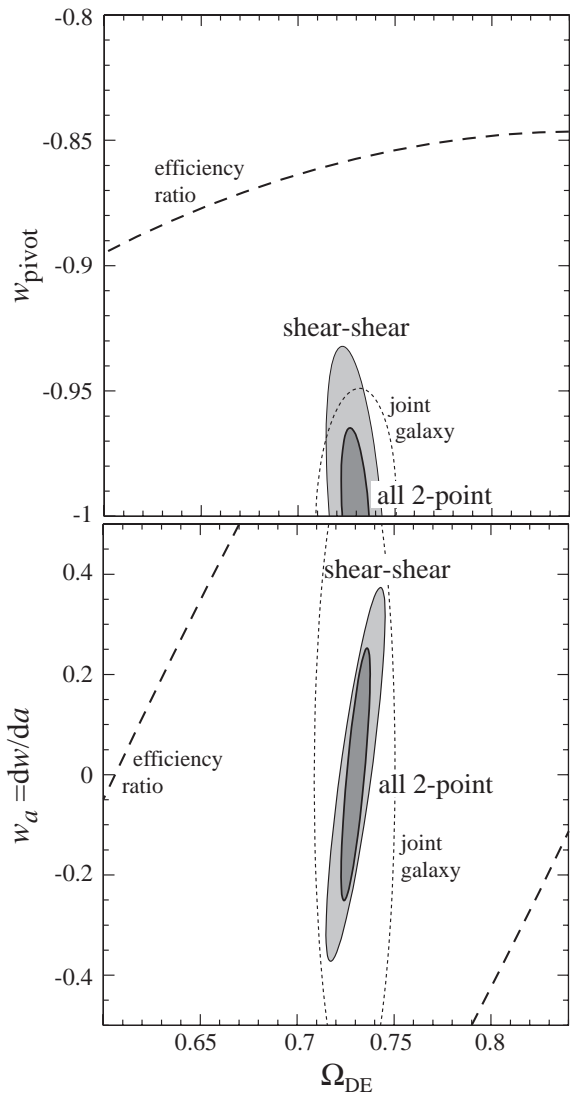}}
\caption{\footnotesize 68\% CL constraints in a three parameter dark energy model
($w_{\rm pivot}$, $w_{a}$, 
$\Omega_{\rm DE}$)  for shear-shear correlations only (light shaded ellipse)
 and all 2-point correlations
(dark shaded ellipses)
with $N_L=10$ and $N_S=5$ and $l_{P}=3000$.   Also shown
for comparison are the efficiency ratio constraints (dashed lines) and joint
galaxy-shear and galaxy-galaxy correlations (dotted lines).
Even after marginalizing over
$5 N_L=50$ halo model parameters, galaxy-shear with galaxy-galaxy power spectra
have comparable constraining power on the dark energy as shear-shear power spectra.
Errors between $w_{\rm pivot}$ and $w_{a}$ are uncorrelated by definition.  The pivot point
$z_{\rm pivot}=0.36$ and is close to the epoch of dark energy domination.
}
\label{fig:summary}
\end{figure}

Given the underlying halo model parameterization, cosmological and halo
model parameters can be jointly fit to the observable power spectra.
Let us
first consider constraints that are based on the galaxy-mass cross-correlation 
alone.  In Fig.~\ref{fig:cross} (dashed lines) we show the cross power
constraints for constant $w_{\rm DE}$-$\Omega_{\rm DE}$ 
and $N_L=10$, $N_S=2$. We marginalize
$5 N_L=50$ halo parameters and vary the maximum $l$ employed in 
Eqn.~(\ref{eqn:Fishercross}) [see Appendix B]. 
Let us focus on the opposite region to the previous section, that of large scales $l < l_P
\sim 10^3$ 
where non-Gaussianity in the fields and inadequacies in the halo model are
minimized.    
Under the halo model prescription, there is sufficient 
information in the power spectra to constrain a 
degenerate combination of $w_{\rm DE}$ and $\Omega_{\rm DE}$.
Note that the errors in the best constrained direction 
are insensitive to the maximum $l_P$ and hence to uncertainties in the
non-Gaussianity of power spectrum errors.   They are also then
less sensitive to inadequacies in the halo modeling that appear
on small scales.

The degeneracy line follows a line of constant linear power spectrum
amplitude and distance at the typical  redshift of
the lenses.   Breaking this degeneracy then depends both on the internal
or external determination of parameters in the halo model 
and its overall validity.  Fortunately, the efficiency ratio information from
small scales is complementary in direction.  
Furthermore, information from the other spectra can included.  
In particular the galaxy-galaxy correlations are more sensitive to the assumptions
of the halo model than the galaxy-mass correlation (see Fig.~\ref{fig:param}).
It may be used to cross check and calibrate the halo model parameters
and potentially extend the parameter space as needed.  

In Fig.~\ref{fig:cross}, we illustrate the utility of a joint analysis.
Adding in galaxy-galaxy power spectra information
out to the same $l_P$ substantially
assists dark energy parameter constraints by effectively acting as
a prior and consistency check on the halo model parameters.
The small interior ellipse illustrates the potential for 
simultaneous determinations
of a constant $w_{\rm DE}$ and $\Omega_{\rm DE}$ by combining this information
with that from the efficiency ratio.

\begin{figure}[tb]
\centerline{\epsfxsize=3.0in\epsffile{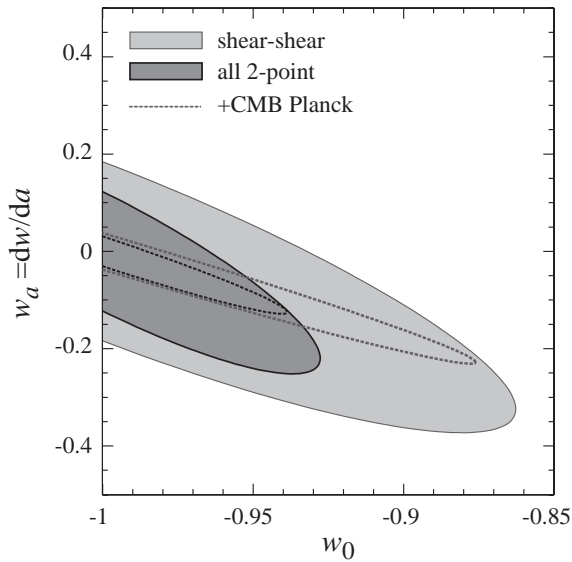}}
\caption{Example of combining external constraints.  The $w_{\rm pivot}$ galaxy-lensing constraints
can be transformed into other dark energy parameterization conventions (here
$\Omega_{\rm DE}$, $w_0$, $w_a$) for comparison and joint studies.  Dashed lines
represent the improvement to the 68\% CL region due to the addition of projected
CMB constraints for the Planck satellite which mainly constrain the angular diameter
distance to recombination.}
\label{fig:w0}
\end{figure}

Just as galaxy-galaxy correlations provide a powerful cross check on halo
model parameter determinations, shear-shear correlations provide a powerful 
cross check on cosmological
parameter determinations.  It is well known that in principle
 shear-shear correlations are an extremely 
powerful probes of the dark energy parameters (e.g. \cite{Hu99b,Hut01}).   

In Fig.~\ref{fig:summary}
we show 
that with shear-shear correlations alone all three dark energy parameters
$w_{\rm pivot}$, $w_a$ and $\Omega_{\rm DE}$ can be simultaneously
measured.  Note that we do not place external prior constraints on $\Omega_{\rm DE}$
unlike the convention commonly found in the literature but that the effect of a prior can be readily
read off these figures since the $\Omega_{\rm DE}$ dimension is shown.
We
have chosen here $N_L=10$ and $N_S=5$.   However shear-shear correlations 
are more susceptible to systematic errors since the cross power
would null out systematics that are not common to both.  They also require  
a modeling of the statistics and their covariance 
in the non-linear regime.  Here we have taken
$l_P=3000$.  

The combined galaxy-shear and galaxy-galaxy power spectra
potentially have  constraining power that is comparable to the shear-shear
spectra.  The combination of all three
reduces the errors by of order $\sqrt{2}$ or more versus shear alone, 
leading to marginalized errors of $\sigma(\Omega_{\rm DE})=0.005$,
$\sigma(w_{\rm pivot})=0.02$ [$\sigma(w_0)=0.05$], 
and $\sigma(w_a)=0.17$.
More importantly
they provide important cross checks against shear systematics on the one
hand and inadequate halo modeling on the other.

For comparison with other probes, it is useful to note that the dark energy
pivot point of the combined power spectra information 
is $z_{\rm pivot}=0.36\approx z_{\rm DE}$, the epoch of dark energy
domination.  Equation~(\ref{eqn:rotation}) 
then maps the errors back into a more conventional description such
as $w_0$ ($a_n=1$).
For example if we add in projected CMB constraints for the Planck satellite
\cite{Planck},
the errors improve to
$\sigma(\Omega_{\rm DE})=0.004$, $\sigma(w_{\rm pivot})=0.01$ [$\sigma(w_0)=0.04$],
$\sigma(w_a)=0.08$ as shown in Fig.~\ref{fig:w0}.  Note that the combined pivot
point shifts to higher redshift.
Here we have employed the sensitivities in \cite{Hu01c} 
amounting to e.g. $\sigma(\ln D_*)=0.002$ 
where $D_*$ is the angular diameter distance to last scattering; note that
current constraints are at the 0.04 level.

\begin{figure}[tb]
\centerline{\epsfxsize=3.2in\epsffile{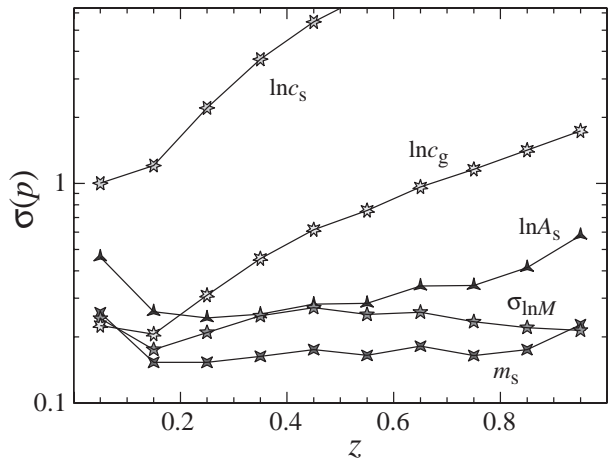}}
\caption{Halo parameter errors as a function of redshift for the $N_L=10$
lens galaxy bins and $N_S=5$ source galaxy bins with
all 2-point information to $l_{P}=3000$ and the efficiency ratio on all scales.
Parameters at different redshifts are largely uncorrelated 
whereas those at the same redshift are highly correlated.   Linear combinations
that control the shape and amplitude of the power spectra are better constrained 
(see text).}
\label{fig:halo}
\end{figure}

Finally, with all three power spectra one can probe the evolution of the
halo parameters and hence aspects of the evolution of the underlying 
galaxy population.  
We show in Fig.~\ref{fig:halo} the resulting
errors on the 5 halo parameters as a function of redshift employing all $2$-point
information out to $l_P=3000$.  The errors
for different redshifts are nearly independent whereas the errors between
the 5 halo parameters at a given redshift are strongly correlated.  The correlation
indicates degeneracies between the parameters as can also be seen in 
Fig.~\ref{fig:param}.  Thus there are combinations of halo parameters that
are better determined than implied by Fig.~\ref{fig:halo}.   These combinations
control the shape and amplitude of the spectra.  For example, the linear
combination that controls the bias and correlation coefficient across the
$0.3 < z <0.4$ bin and at scales central to the constraint, e.g. $k=0.3 h$ Mpc$^{-1}$
are separately constrained at the level of $\sigma(\ln b) = 0.01$ and $\sigma(\ln R) = 0.004$
(fiducial values: $b=2.2$ and $R=1.2$).
Constraints on the correlation coefficient mainly reflect its insensitivity to halo 
parameters on large scales; they weaken in the deeply non-linear regime.
Constraints on the bias as a function of scale remain at the percent level from the
linear to well into the non-linear regime.

\section{Discussion}
\label{sec:discussion}

The joint analysis of galaxy clustering and lensing data from next-generation
surveys offers unique opportunities to simultaneously determine the
evolution of clustering in the dark matter and galaxies. These surveys
are expected to provide multi-color catalogs of galaxies with well 
characterized photometric redshifts well beyond $z=1$. 
We have shown here that with redshift information, even at the 2-point
or power spectrum level, there is enough information in the galaxy-shear
correlation, especially when combined with galaxy-galaxy correlations,
to jointly solve for a model with $\sim 50$
parameters to describe the galaxy evolution and $3$ dark energy parameters.
We have conservatively allowed the galaxy parameters to vary independently 
as a function of lens redshift and marginalized the associated parameters when quoting
dark energy constraints.  

The dark energy parameter determinations are statistically
competitive with the shear-shear correlations and should be more
robust to systematic errors in the shear determinations.  They furthermore
can provide a better redshift localization of dark energy effects given the
broad lensing kernel of the shear.

These determinations are assisted by two relatively robust features in the
halo model: on large scales, the combination of galaxy-mass correlations
and galaxy-galaxy correlations may be used to determine the underlying
mass-mass correlations in a manner that is only weakly sensitive to
halo model assumptions; on small scales the ratio of galaxy-shear correlations
at different source redshifts yields distance ratio information that
is only weakly dependent on the 
evolution of the galaxy-mass correlation.

Equally important, by combining galaxy-galaxy and galaxy-shear clustering one
can determine whether the halo model employed here suffices as a description
of the relationship between galaxies and dark matter halos.  
Consistency of the halo determinations can also be checked by selecting
lens galaxies with multiple cuts on luminosity or rarity.  Consistency of the
dark energy determinations can be checked against the shear-shear correlations
which depend only on the mass spectrum.  
A further extension to include contributions from the many galaxy-shear bispectra 
would
also improve parameter accuracies and provide cross-checks
\cite{Hui99,TakJai03}.

Halo model
parameter determinations will be valuable in
testing models of galaxy formation (e.g.~\cite{Beretal03,Scr03,Zehetal03}). 
Galaxy parameters measured from
surveys as a function of galaxy type, luminosity and redshift can 
be compared with simulations and semi-analytic
model predictions of the same parameters.  
Although these studies may show that the our halo model requires 
further modification and extension, we believe that the prospects are
bright for a joint solution of galaxy and dark matter clustering. 

\noindent{\it Acknowledgments:}  We thank G. Bernstein, D. Holz, 
A. Kravtsov, M. Takada, A. Tasitsiomi and
especially E. Sheldon for many useful discussions.  W.H. is supported
by NASA NAG5-10840, the DOE, and the Packard Foundation. B.J. is supported by
NASA NAG5-10923 and the Keck Foundation.

\appendix

\section{Cosmological Power Spectrum}
\label{sec:linear}

The linear mass power spectrum $P(k)$
which underlies all observable power spectra
depends only on well-motivated cosmological parameters 
in the context of the successful $\Lambda$CDM cosmology.  
In this section, we describe its parameterization in terms of
the initial conditions \S\ref{sec:initial} and evolution
\S\ref{sec:evolution}.

\subsection{Initial Conditions}
\label{sec:initial}

We begin by assuming that massive neutrinos make 
a negligible contribution to the matter density.
The shape of the mass power spectrum is then specified by the baryon 
density $\Omega_{b}h^{2}$, the dark matter density $\Omega_{m}h^{2}$,
and the spectrum of initial curvature fluctuations $\zeta$
\begin{equation}
\Delta_{\zeta}^{2}=\delta_{\zeta}^{2} \left( {k \over k_{0}}\right)^{n-1}\,,
\label{eqn:initialconditions}
\end{equation}
where $k_{0}=0.05$ Mpc$^{-1}$ is the normalization scale.  
We take fiducial values for these parameters that are consistent with
the WMAP determinations: $\Omega_{b}h^{2}=0.024$, 
$\Omega_{m}h^{2}=0.14$, $n=1$ and $\delta_{\zeta}=5.07 \times 10^{-5}$
\cite{Speetal03}.
The current uncertainties in these parameters are at the 10\% level or
better.  Note that $\delta_\zeta$ is related to the WMAP normalization 
parameter by
\begin{equation}
A = (1.84\delta_\zeta \times 10^4)^2\,, 
\end{equation}
and current and future uncertainties in this parameter are expected
to be dominated by uncertainties in the Thomson 
optical depth to reionization $\tau$, i.e.
\begin{equation}
\delta_\zeta \approx 5.07 e^{-(0.17-\tau)} \times 10^{-5}. 
\end{equation}
Thus the power spectrum as a function of $k$ in Mpc$^{-1}$ (not
$h$ Mpc$^{-1}$) in the matter dominated regime can be
considered as largely known.

\subsection{Evolution}
\label{sec:evolution}

The shape of this initial power spectrum does not change during
dark energy domination 
on scales  below the sound horizon of the dark energy.
The amplitude of the linear power spectrum
depends on the initial normalization
$\delta_{\zeta}$ and the ``growth function", here the 
decay rate of potential fluctuations $G$
\begin{equation}
P(k,z) =\left[ {1 \over {1+z}} {G(z) \over G_0} \right]^2 P(k,0) \,,
\end{equation}
where $G_0\equiv G(z=0)$ and
we assume that all relevant scales are sufficiently below the 
maximal sound horizon of the dark energy.   $P(k,0)$ can  be
evaluated from any one of a number of Einstein-Boltzmann
codes.

The normalization of the linear power spectrum today is conventionally given
at a scale of $r=8 h^{-1}$Mpc
\begin{eqnarray}
\sigma_8^2 &\equiv& \int {d^3 k \over (2\pi)^3} P(k,0) W_\sigma^2(kr)\nonumber\\
\sigma_{8} &\approx& {\delta_{\zeta} \over 5.59\times10^{-5}} 
\left( { \Omega_{b}h^{2} \over 0.024} \right)^{-1/3}
\left( { \Omega_{m}h^{2} \over 0.14} \right)^{0.563}\nonumber\\
&&\times(3.123h)^{(n-1)/2} \left( { h \over 0.72} \right)^{0.693}  
{G_0 \over 0.76}\,,
\label{eqn:sigma8}
\end{eqnarray}
where $W_\sigma(x)= 3x^{-3}(\sin x - x \cos x)$ is the Fourier transform of
a top hat window.   The approximation in Eqn.~(\ref{eqn:sigma8}) 
is valid to the 1\% level for individual
variations of the parameters in the regime
$0.019 \le \Omega_{b}h^{2} \le 0.03$,
$0.11 \le \Omega_{m}h^{2} \le 0.18$, $0.7  \le n \le 1.3$,
$0.5 \le h \le 1$ which more than span the current observational errors.
Note that because the normalization is given in $h^{-1}$ Mpc, there is
a strong scaling with the Hubble constant.  This scaling actually
assists dark energy determinations since in a flat universe $\Omega_{\rm DE}
= 1-\Omega_m$ and 
precise measurements of $\Omega_m h^2$ makes 
$h$ depend on $\Omega_{\rm DE}$ only.  A measurement of the Hubble constant
is a measurement of the dark energy density.  Likewise a measurement
of $\sigma_8$ is a measurement of a specific combination of dark energy
parameters.  

We will hereafter limit ourselves to flat
universes but correspondingly {\it neglect} the dark energy 
information from the CMB unless otherwise specified.
The angular diameter distance to recombination $D_*$ has been
measured to $\sigma(\ln D_*)=0.04$ \cite{Speetal03} and this constraint will continue to 
improve with better measurements of $\Omega_{m}h^{2}$ from
the peaks as (see e.g. \cite{HuFukZalTeg00})
\begin{equation} 
 \sigma(\ln D_*) \approx
{1 \over 4} \sigma(\ln \Omega_m h^2) \,.
\end{equation}
The rationale behind dropping this constraint
is that this measurement will be used in conjunction with
galaxy and lensing constraints to eliminate any small curvature 
contribution that might exist.  In a flat universe, $D_{*}$ closely follows $G_0$ 
in its dark energy dependence and so 
may be used as a powerful consistency test for the absence of spatial 
curvature.   We discuss this issue further in \S \ref{sec:full}.

The growth function $G$ then depends only on the dark energy density 
$\Omega_{\de}(a)=8\pi G \rho_{\rm DE}/3H^2$ 
and equation of
state $w(a)=p_{\rm DE}/\rho_{\rm DE}$ through the equation (e.g. \cite{Hu01c})
\begin{eqnarray}
\frac{d^2 G}{d \ln a^2}  &+&
\left[ \frac{5}{2} - \frac{3}{2} w(a) \Omega_{\de}(a) \right]
\frac{d G}{d \ln a}   \nonumber\\
&+&
\frac{3}{2}[1-w(a)]\Omega_{\de}(a) G =0\,,
\label{eqn:growth}
\end{eqnarray}
where the initial conditions are
$G=1$ and $dG/d\ln a=0$ at an epoch well before dark
energy domination  $z\gg z_{\rm DE}$,
\begin{equation}
1+z_{\rm DE} = a_{\rm DE}^{-1} \approx 
\left( {\Omega_{\rm DE}\over {1-\Omega_{\rm DE}}} \right)^{-1/3 w_0}\,,
\label{eqn:dedom}
\end{equation}
where $\Omega_{\rm DE}\equiv \Omega_{\rm DE}(z=0)$ and $w_0\equiv w(z=0)$.

Given a constant equation of state, $G_0$ follows the 
approximate form
\begin{eqnarray}
G_0 &\approx& 0.76 \left( {\Omega_m \over 0.27} \right)^{0.236} 
	F[\Omega_{\rm DE}^{4/3} (1+w_{\rm DE})]\,, \nonumber\\
F(x) &=& (1+0.498 x + 4.88 x^3)^{-1}\,,
\label{eqn:growthfunction}
\end{eqnarray}
where the approximation holds to $\sim 1\%$ for separate variations of
$-1< w_{\rm DE}< -1/2$ and $0.12 < \Omega_m < 0.5$.  
Here and throughout we denote a dark energy parameterization for
which $w$ is constant with $w_{\rm DE}=w$.
Note the
fairly strong scaling of $G_0$ with $w_{\rm DE}$ around the fiducial model
$\Delta G_0/G_0 \approx -0.33 \Delta w_{\rm DE}$.

A dynamical form of dark energy is unlikely to possess
a strictly constant equation of state.   
Since it only has observable consequences at 
$z \simlt z_{\rm DE}$, it is convenient to describe the function with
its first order Taylor expansion.
We therefore choose an equation of state parameterization
\begin{equation}
w(a) = w_{n} + (a_{n}- a) w_a\,,
\label{eqn:wn}
\end{equation}
with the expansion around some epoch $a_n$; this generalizes
the form employed in \cite{Lin03} where $a_{n}=1$.

We instead choose $a_{n}=a_{\rm pivot}$ such that 
the errors in $w_{\rm pivot}$ and $w_a$ for a given observable are uncorrelated 
$C_{w_{\rm pivot} w_a}=0$.  The pivot point $a_{\rm pivot}$ can be derived
from a general representation via the transformation
\begin{equation}
C_{\mu\nu}' = \sum_{\alpha\beta} 
{\partial p_\alpha' \over \partial p_\mu} 
C_{\alpha\beta}
{\partial p_\beta' \over \partial p_\nu} \,.
\label{eqn:rotation}
\end{equation}
For the transformation to the pivot representation \cite{EisHuTeg99b}
\begin{eqnarray}
{\partial w_{\rm pivot} \over \partial w_{n}} = 1, 
\qquad
{\partial w_{\rm pivot}
 \over \partial w_a} = a_{n} - a_{\rm pivot}\,, 
 \label{eqn:wpivot}
\end{eqnarray}
from which it follows that the errors decorrelate for a shift
in the pivot of
\begin{equation}
{a_{n} - a_{\rm pivot}} = -{C_{w_{n}w_a}\over 
C_{w_a w_a}}\,.
\end{equation}
Moreover the resulting errors on $w_{\rm pivot}$ are then equal to
those on $w_{\rm DE}$, a constant $w$.
The pivot redshift is therefore also the redshift 
at which $w$ is best constrained.

The drawback to choosing the pivot redshift 
is that it is specific to the observable, survey and dark energy model.  
As the redshift where the dark energy evolution is best constrained,
the pivot point for growth measurements
tends to coincide roughly with $z_n\approx 0.4$ or
$z_{\rm DE}$ in the fiducial $\Lambda$ model.
In fact, further
taking $w_{\rm DE} \rightarrow w(a_{\rm DE})$ in Eqn.~(\ref{eqn:growthfunction})
yields an approximation to the growth function that is good to several
percent across a wide range of $w_a$.
Choosing this standard normalization epoch then provides 
the benefits of the pivot point without the drawbacks.    
Although we will employ the pivot redshift for the galaxy-lensing
study below, it is 
sufficiently close to $z_{\rm DE}$ that
it may be interpreted in this manner.  Note that the pivot point for 
distance-based dark energy measurements can be at even higher redshift
$z_{\rm DE} \simlt z \simlt 1$ so that it becomes even more important to
choose a $z_{n} \ne 0$ for the characterization of constraints.

In summary, our linear matter power spectrum is specified by $7$ 
parameters: 4 that are already well constrained by the CMB
$\delta_\zeta (=5.07\times 10^{-5})$, 
$n (=1)$, 
$\Omega_bh^2 (=0.024)$, 
$\Omega_m h^2 (=0.14)$, 
and 3 dark energy
parameters 
$\Omega_{\rm DE} (=0.73)$, 
$w_{\rm pivot} (=-1)$ (or $w_0$), 
$w_a (=0)$ which galaxy-lensing
power spectra can help constrain.  Parameter values of the fiducial
cosmology are given in parentheses.
Given a linear power spectrum and cosmology, cosmological simulations
can accurately predict the fully non-linear mass power spectrum and hence
the shear-shear 2-point correlations.  
On the other hand, the power spectra involving galaxies
will require some semi-analytic modeling for the foreseeable future.
We now turn to a halo model for the parameterization of the
underlying relationship between the observable power spectra
and the linear theory predictions.

\begin{figure}[tb]
\centerline{\epsfxsize=3.2in\epsffile{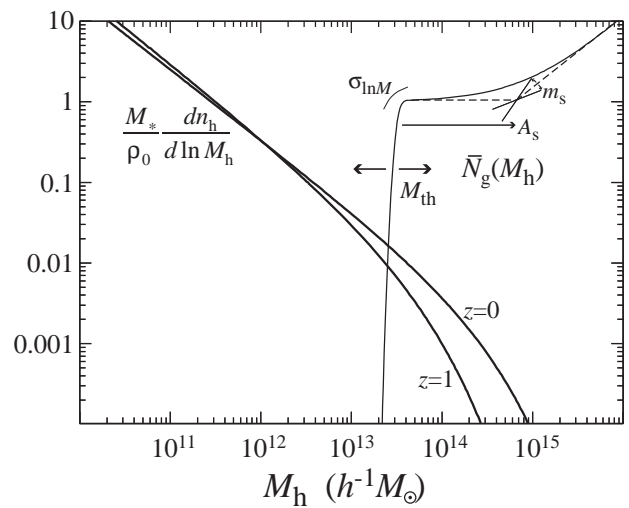}}
\caption{\footnotesize Host halo mass function $dn_{h}/d\ln M_{h}$ and halo occupation distribution
function $\bar N_g(M_{h})$.    Three halo parameters control the shape of the
distribution: $A_{s}$ the
satellite-host 
normalization or crossing point, $m_{s}$ the satellite slope (which pivots $\bar N_s$ 
about the crossing point), and $\sigma_{\ln M}$ the scatter in
the mass observable relation.
Galaxies selected by a flux limit are matched in number density 
by adjusting the threshold mass $M_{\rm th}$, here illustrated
for $M_{\rm th}=10^{13.5} h^{-1} M_{\odot}$.
Here and throughout the fiducial model is a flat $\Lambda$CDM model with cosmological
parameters
$\Omega_{m}=0.27$, $h=0.72$, $n_{s}=1$, $\delta_{\zeta}=5.07 \times 10^{-5}$
($\sigma_{8}=0.91$) and $\Omega_{b}h^{2}=0.024$
and halo parameters $A_{s}=30$, $m_{s}=1$ and $\sigma_{\ln M}=0.1$.}
\label{fig:massfn}
\end{figure}

\section{Halo Model}
\label{sec:halo}

To study the information contained in the lensing and galaxy two
point observables, we require a model for the underlying three dimensional
galaxy and mass density power spectra.   Recent work on comparing simulations to
galaxy clustering data have shown that to first approximation, galaxies 
selected on luminosity are assigned to a mass-based selection of
dark matter (sub)halos of the same spatial number density
\cite{Kraetal03}.   This mapping avoids the traditional problem of defining explicit halo mass-luminosity relationships.
In \S \ref{sec:satellite}, we build this underlying ansatz into a halo description
of the galaxies and mass (see \cite{CooShe02} for a review) and obtain a parameterized model for their
joint power spectra in \S \ref{sec:halopower}.  
We then describe the phenomenology of the fiducial halo model in \S\ref{sec:phenomenology}
and study the sensitivity of the power spectra to variations in the halo model
and cosmological parameters in \S \ref{sec:sensitivity}.

\subsection{Host and Satellites}
\label{sec:satellite}

Under the halo model, the power spectra are described by the abundance, clustering,
density profile, substructure and galaxy occupation of dark matter halos.  
For the comoving abundance of host halos, we take 
the mass function  \cite{SheTor99}
\begin{eqnarray}
{d n_h \over d\ln M} = {\rho_{0} \over M} f(\nu) {d\nu \over d\ln M}\,,
\label{eqn:massfn}
\end{eqnarray}
where $\rho_{0}$ is the matter density today $\rho_{0}= \rho_{m}(z=0)$,
$\nu = \delta_c/\sigma$ 
\begin{eqnarray}
f(\nu) = A\sqrt{{2 \over \pi} a\nu^2 } [1+(a\nu^2)^{-p}] \exp[-a\nu^2/2]\,.
\end{eqnarray}
Here $\sigma(M;z)$ is the rms of the linear density field smoothed with
a top hat of a radius that encloses the mass $M$.  We choose
$\delta_c=1.69$, $a=0.75$, $p=0.3$, and $A$ such that $\int d\nu f(\nu)=1$.
The mass function of the fiducial model is shown in Fig.~\ref{fig:massfn}
bracketing the redshifts of interest for lensing ($0 < z < 1$).

The halo clustering is then given by the peak-background split 
as \cite{Kai84b,MoWhi96}
\begin{equation}
b(M) = 1 + {a \nu^2 -1 \over \delta_c}
         + { 2 p \over \delta_c [ 1 + (a \nu^2)^p]}\,.
\label{eqn:bias}
\end{equation}
For the halo density profile,
we take the NFW form \cite{NavFreWhi97}
\begin{equation}
\rho(r,M,c) = {\Delta_v \rho_0\over 3} {c^3  \over \ln(1+c) - c/(1+c)} {1 \over R c(1+ R c)^2}\,,
\end{equation}
where $R=r/r_v$ with the virial radius
\begin{equation}
r_v = \left( {3 M_v \over 4\pi \rho_{0} }\right)^{1/3}\,.
\label{eqn:virialradius}
\end{equation}
We take the concentration \cite{Buletal01}
\begin{equation}
c(M_v) = {9 \over {1+z} } \left( {M_v \over M_*} \right)^{-0.13}\,,
\label{eqn:concentration}
\end{equation}
where $M_{*}$ is defined by  $\sigma(M_*;z=0)=\delta_c$. 
We convert between the halo mass $M$, assumed to be defined at an overdensity of 
180 times the mean density, and the virial mass $M_v$ defined at an overdensity 
\cite{BryNor98}
\begin{equation}
\Delta_v(z) = {18\pi^2 + 82 \Omega_{\rm DE}(z) - 39 [\Omega_{\rm DE}(z)]^2 \over 1+
\Omega_{\rm DE}(z)}\,,
\end{equation}
using the NFW profile (see e.g. \cite{HuKra02}).

\begin{figure}[tb]
\centerline{\epsfxsize=3.2in\epsffile{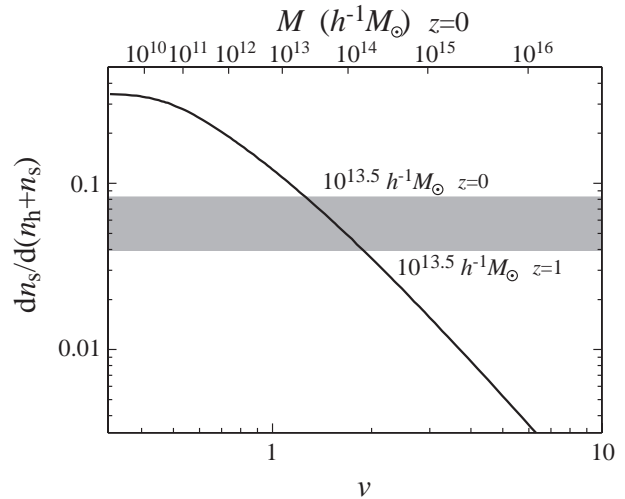}}
\caption{\footnotesize Satellite fraction $dn_{s}/d(n_{h}+n_{s})$.
The satellite fraction at a given mass 
for a given halo occupation distribution
 $N(M_{h})$ is largely a function of the peak height threshold $\nu$ in
Eqn.~(\ref{eqn:massfn}) as rare objects are unlikely
 to be satellites.  The conversion to mass at $z=0$ is given in the upper axis.   
The
 range of fractions for $M=10^{13.5} h^{-1} M_{\odot}$ and $0 < z < 1$ is shown shaded.}
\label{fig:satellite}
\end{figure}

The profile enters into the power spectra through its normalized Fourier
transform
\begin{equation}
y(k,M,c) = {1 \over M} \int_0^{r_v} dr
4\pi r^2 \rho(r,M,c) {\sin (kr) \over kr} \,.
\end{equation}
Although there are current uncertainties in these descriptions of the
halo mass function,
bias and concentration, we choose {\it not} to associate free parameters with these
{\it dark matter} clustering properties.  The characterization of
these  properties and more importantly the mass power spectrum
itself can in principle be fixed by better simulations.   
Similarly for the association with galaxies, 
since we will be matching number densities of objects,  the halo mass
definition employed here may be replaced with any variable that defines the
selection of objects in the simulations and is a monotonic function of
galaxy luminosity on average.

The halo model predicts the galaxy and galaxy-mass power 
spectra under an assumption
for the statistics of the occupation of the host halo by galaxies \cite{Sel00,SheJai03}. 
Since
each galaxy also carries its own dark matter halo we will hereafter
distinguish between the host halo of mass $M_h$ and its satellite
halos of mass $M_s$.  For simplicity we take the satellites to also
have NFW profiles but with an adjustable concentration; 
a more sophisticated model would account for the change in the
functional form due to truncation from tidal stripping and trends in
mass.  

\begin{figure}[tb]
\centerline{\epsfxsize=3.2in\epsffile{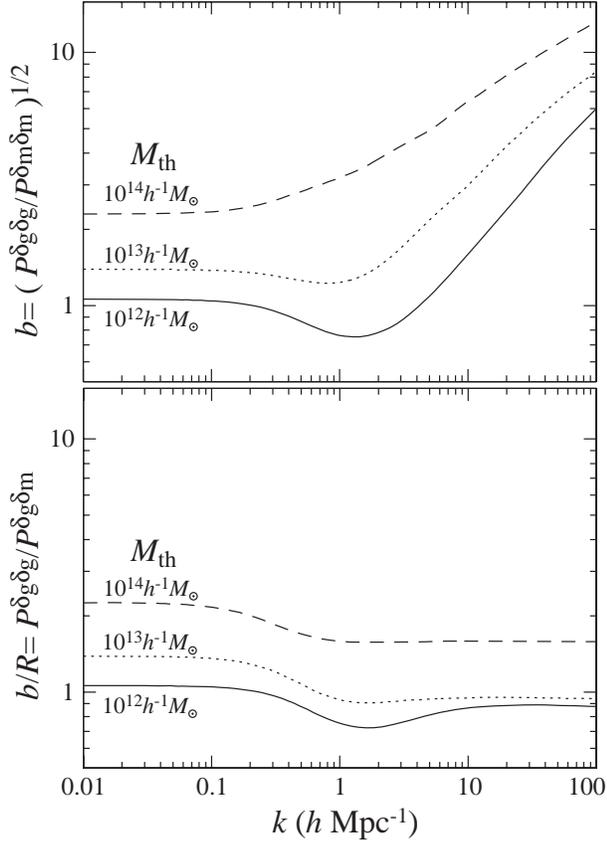}}
\caption{Relative galaxy-galaxy and galaxy-mass power spectra.  In the linear regime, $k < 0.1 h$ Mpc$^{-1}$, 
both $b=(P^{\dgg}/P^{\dmm})^{1/2}$ and $b/R = P^{\dgg}/P^{\dgm}$
return the linear bias which increases as the objects become rarer.  In the non-linear regime,
$P^{\dgg}$ and $P^{\dgm}$ continue to track each other and exceed the mass power spectrum.}
\label{fig:bias}
\end{figure}

We begin by taking a simulation-motivated universal form for the
number of dark matter satellites in the host halo as
a function of their mass ratio \cite{TorDiaSye98,Ghietal00}
\begin{equation}
{d \bar N_s \over d\ln M_s} = m_s \left( { M_h \over A_s M_s } \right)^{m_s}
\label{eqn:satmass}
\end{equation} 
for $M_h/(100A_s) \le  M_s \le  M_h $
where simulations suggest that $A_s \approx 30$ and
$m_s \approx 1$ \cite{Kraetal03}.  The cutoff to low masses prevents a mild logarithmic
divergence in the total mass; this arbitrary cutoff only affects
the mass power spectrum at unobservably small scales.

We next associate galaxies with these satellites.
The total number of satellite galaxies above a given threshold mass $M_{\rm th}$ then becomes
\begin{eqnarray}
\bar N_s &=& \int_{M_{\rm th}}^{\infty} {dM_s \over M_s} {d \bar N_s \over d\ln M_s} 
\nonumber\\
	 &=& \left( {M_h \over A_s M_{\rm th}} \right)^{m_s}\,.
	 \label{eqn:satellitenumber}
\end{eqnarray}
To this population of satellite galaxies we add the central galaxy associated with 
the host halo itself
to obtain the total number of galaxies above the threshold (see Fig.~\ref{fig:massfn})
\begin{equation}
N_{g}(M_h;M_{\rm th}) = N_h + N_s\,.
\label{eqn:hostsatellitesplit}
\end{equation}

These two pieces have different statistical properties.
The central
galaxy may be either above or below threshold and hence
either occupied or unoccupied leading to 
$\langle N_h^p \rangle = \langle N_h \rangle = \bar N_h $.  We model the
average value as a step function at some limiting mass $M_{\rm th}$ 
smoothed by a Gaussian in $\ln M_h$ to reflect scatter in the conversion
of a magnitude limit to a mass limit
\begin{equation}
\bar N_h = {1 \over 2}{\rm Erfc}\left( {\ln M_{\rm th}/M_h}
\over \sqrt{2} \sigma_{\ln M} \right)\,.
\end{equation}
The dark matter satellites follow a Poisson distribution with
$\langle N_s^2 \rangle = \bar N_s^2 + \bar N_s$ \cite{Kraetal03}.

Given a fixed shape for $\bar N_g$, the threshold mass $M_{\rm th}$ is not a
free parameter.   Rather it is fixed by matching the space density of
galaxies to the space density inferred by the  observed number counts
\begin{eqnarray}
\bar n_{V}(z;M_{\rm th}) &= &\int_{0}^{\infty} {d M_{h} \over M_{h}} \bar N(M_{h};M_{\rm th}) {d n_{h}\over d\ln M_{h}}\nonumber\\
& =&  \bar n_A W_g(z)\,.
\end{eqnarray}
Note that by choosing rare objects, the implied $M_{\rm th}>M_*$
and so the population will be dominated by host galaxies rather than
satellite galaxies (see Fig.~\ref{fig:massfn}).  
This fact will be useful in minimizing the uncertainties and systematic
errors associated with the model for $\bar N_s$ and the satellite
profiles, e.g. the mass scaling of the concentration parameters.  
The satellite contribution can be quantified by considering 
the satellite mass function 
\begin{eqnarray}
{d n_s \over d\ln M_s} &=& \int_{M_s}^{\infty} {d M_h\over M_h}
{d \bar N_s \over d\ln M_s} {d n_h \over d\ln M_h}\nonumber\\
& =&  \int_{M_s}^{\infty} {d M_h\over M_h}
m_s \left( { M_h \over A_s M_s } \right)^{m_s}
{d n_h \over d\ln M_h}
\label{eqn:satellitemass}
\end{eqnarray}
and comparing it to the host halo mass function itself
\begin{equation}
{dn_{s} \over  d(n_{h}+n_{s})} \equiv
{ d n_{s} \over d\ln M } \left[
{ d (n_{s}+n_{h}) \over d\ln M } \right]^{-1}\,.
\end{equation}
This ratio depends mainly on the peak
height threshold $\nu$ for a given form for the shape of $\bar N_g(M_h)$  
(see Fig.~\ref{fig:satellite}).  With our fiducial choice, it saturates
at the low mass end at $\sim 30\%$. With a selection of rare objects,
e.g. corresponding to $M_{\rm th} \approx
10^{13.5} h^{-1} M_\odot$ in the fiducial model, the satellite fraction
can be less than $10\%$ at all redshifts.

 \begin{figure}[tb]
\centerline{\epsfxsize=3.2in\epsffile{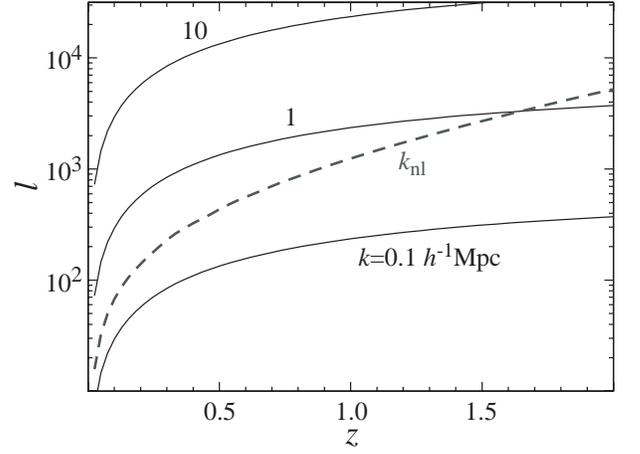}}
\caption{\footnotesize Correspondence of spatial wavenumber $k$ ($h^{-1}$ Mpc) to
angular  wavenumber $l$ as a function of redshift $z$ given by the Limber approximation
in the fiducial cosmology.
Also shown is the projected non-linear scale $k_{\rm nl}$ where $k^3 P^{\dmm}/2\pi^2=1$.
Note that $l\sim 10^3$ corresponds to the mildly non-linear regime for the
redshifts in question.}
\label{fig:ktol}
\end{figure}

\subsection{Power Spectra}
\label{sec:halopower}

The power spectra are defined by the host and satellite distributions as
\begin{eqnarray}
P^{\delta_m \delta_m}(k) &=& I_{1m}^2(k) P(k) + I_{2 m}(k)\,,\nonumber\\
P^{\delta_g \delta_g}(k) &=& I_{1g}^2(k) P(k) + I_{2 g}(k)\,,\nonumber\\
P^{\delta_g \delta_m}(k) &=& I_{1g}(k) I_{1m}(k) P(k) + I_{2 c}(k)\,.
\end{eqnarray}
The indices $1$ and $2$ refer to the number of points in a single halo.
The first piece then represents two points
in separate host halos correlated by the linear power spectrum $P(k)$
\begin{eqnarray}
I_{1m} &=& \int {d M_h \over M_h} \left( { M_h \over \rho_{0} } \right) 
		{dn_h \over d\ln M_h}
		b(M_h) y_h\,, \\
I_{1g} &=& {1 \over \bar n_V} \int {d M_h  \over M_h}
[\bar N_h  + \bar N_s y_g]{dn_h \over d\ln M_h} 
		b(M_h)\,, \nonumber 
\end{eqnarray}
and the second piece is the contribution from two points within
a parent halo including its satellite contributions
\begin{eqnarray}
I_{2m} &=& \int {d M \over M} \left( { M \over \rho_{0} } \right)^2
	\left[ {d n_h\over d\ln M} y^2_h 
	+ {d n_s \over d\ln M} y^2_s \right]\,, \nonumber\\
I_{2g} &=& {1 \over \bar n_V^2} \int {d M \over M}
	[\bar N_s^2 y_g^2 + 2 \bar N_h \bar N_s y_g]
	{d n_h\over d\ln M} \,, \nonumber\\
I_{2c} &=& {1 \over \bar n_V} \int {d M \over M}
\left( {M \over \rho_{0}} \right)
\Big[ {d n_h\over d\ln M} 
	(\bar N_h y_h + \bar N_s y_h y_g)\nonumber\\
      && +  {d n_s\over d\ln M} H(M-M_{\rm th}) y_s \Big]  \,,
\label{eqn:singlehalo}
\end{eqnarray}
Here we have employed the shorthand convention $y_{h,s} \equiv
y(k,c_{h,s},M_{h,s})$, $y_g=y(k,c_g,M_h)$   
and the step function $H(x)=1$ for $x\ge 0$ and $H(x)=0$ for $x\le 0$.
The effect of satellite galaxies occupying satellite halos takes
the same form as the central galaxy occupying the host halo \cite{GuzSel02}
 except that
we have neglected the small effect of smoothing around the threshold mass.
In this description, the 
satellite-satellite mass correlation
and satellite-halo mass correlation are implicitly included as the replacement
of mass lost to subhalos in the term involving the parent profiles $y_h$ in
Eqn.~(\ref{eqn:singlehalo}).

\begin{figure}[tb]
\centerline{\epsfxsize=3.2in\epsffile{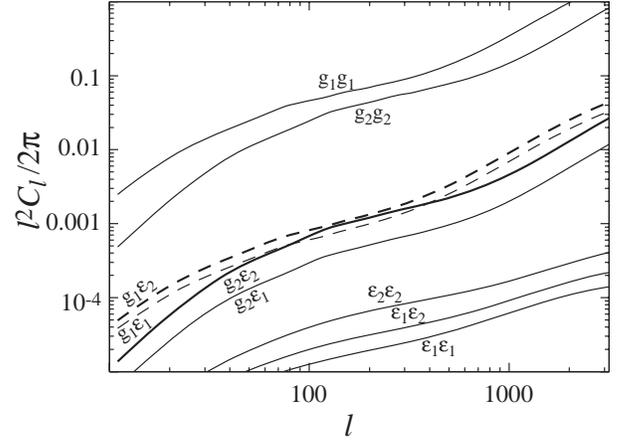}}
\caption{Angular power spectra in the fiducial halo and cosmological
model for two galaxy ($g_1: 0<z<0.5$; $g_2: 0.5<z<1$; 
$M_{\rm th}=10^{13.5} h^{-1} M_\odot$)
and source bins 
($\epsilon_1: z<1.5$; $\epsilon_2: z>1.5$; $z_{{\rm med};S}=1.5$).  Note the
inflection at $l \sim$ few $\times 10^{2}$ and the large number of cross
spectra $(N_{L} N_{S}=4)$.}
\label{fig:cl}
\end{figure}

\subsection{Fiducial Model}
\label{sec:phenomenology}

Five functions of redshift specify our halo model, the satellite-host normalization $A_s$,
the slope of the satellite mass function and occupation distribution $m_s$, the
scatter in the mass observable relation $\sigma_{\ln M}$, the concentration of the
satellites in the host halo $c_g$ and the concentration of the mass profiles
of the satellites $c_s$.  We assume that the latter two follow the mass scaling
of isolated halos in Eqn.~(\ref{eqn:concentration}) but have an arbitrary normalization.
We further represent the functions by a set of $5N_L$ parameters that specify 
their values at the redshifts of the $N_L$ lens redshift bins.

\begin{figure}[tb]
\centerline{\epsfxsize=3.5in\epsffile{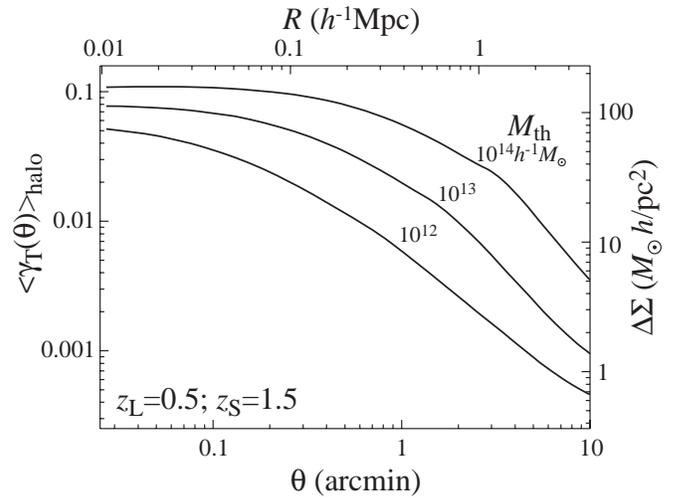}}
\caption{Tangential shear and surface mass density difference in the fiducial
halo and cosmological model for several choices of the mass threshold.  
Here we take a delta function distribution of lenses ($z_L=0.5$)
and sources ($z_S=1.5$).  Note the turnover on small scales 
($0.5-1'$) and inflection near $10'$ causing excess signal at large angles.}
\label{fig:gamt}
\end{figure}

The values of these parameters will in the future be determined by fitting
to the joint power spectra.  However, to study the potential of these future
data sets we must specify their values in the fiducial model.  The sensitivity
of observables to variations in their values around this model is then quantified
through the Fisher matrices of \S \ref{sec:fisher}.

We choose fiducial values that are roughly consistent
with low redshift measurements from the SDSS lensing data \cite{Sheetal03}
and $N$-body simulations: $A_s=30$, $m_s=1$, $\sigma_{\ln M}=0.1$ for all
redshift bins.  These
three parameters define the shape of $\bar N_g$ shown in Fig.~\ref{fig:massfn}.
In our fiducial model the concentration parameters  $c_s=c_{g}=c_{h}$.

In Fig.~\ref{fig:bias}, we show the galaxy-galaxy and galaxy-mass power spectra
with the fiducial halo and cosmological parameters.  
They are shown relative to the mass-mass spectrum 
at $z=0$ for halo abundances
corresponding to several different mass thresholds in the fiducial
$\Lambda$CDM cosmology (see \S \ref{sec:linear}).  In the linear regime $k <  0.1 h$ Mpc$^{-1}$
the ratios
\begin{equation}
b(k) \equiv \left( { P^{\dgg} \over P^{\dmm} } \right)^{1/2}
\label{eqn:biask}
\end{equation}
and
\begin{equation}
{b(k) \over R(k)} \equiv {P^{\dgg}\over P^{\dgm}}
\end{equation}
both return the constant linear bias of the objects.  In other words, in the linear
regime the correlation coefficient
\begin{equation}
R(k) \equiv {P^{\dgm}\over (P^{\dgg}P^{\dmm})^{1/2}} \approx 1
\end{equation}
independent of the population or halo model parameters.
In this regime, a measurement of $P^{\dgm}$ and $P^{\dgg}$ is a measurement
of the mass power spectrum $P$ and hence in combination can be used
to study the dark-energy dependent growth rate.  
The linear bias
increases with the mass threshold $M_{\rm th}$ or more generally
with the rarity of the objects.   

\begin{figure}[tb]
\centerline{\epsfxsize=3.2in\epsffile{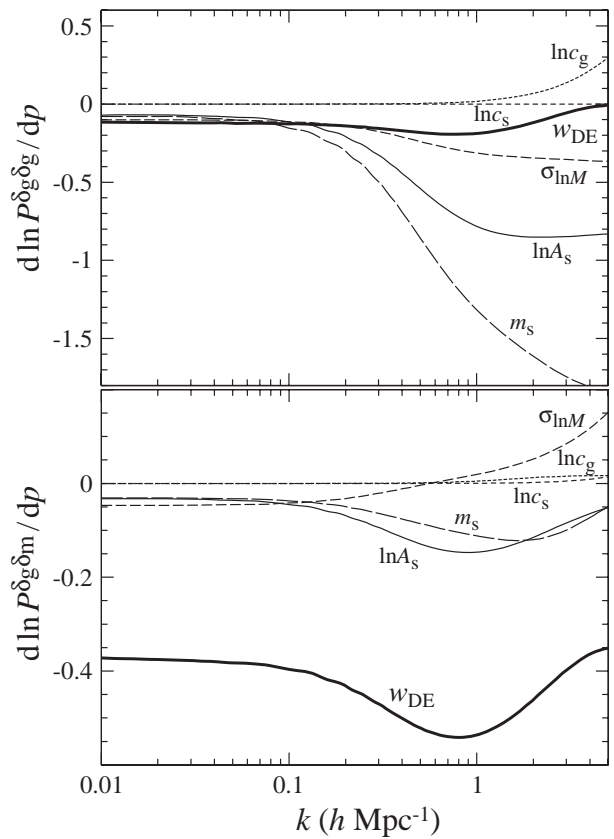}}
\caption{\footnotesize Halo parameter sensitivity compared with dark energy sensitivity 
of galaxy-galaxy and galaxy-mass power spectra.
In the non-linear regime, galaxy-galaxy power spectra become highly sensitive
to most halo parameters.  Galaxy-galaxy power spectra are insensitive to dark
energy effects on the growth rate since a lowering of the amplitude is compensated
by an increase in bias due to the increased rarity of the objects.  Conversely the
galaxy-mass power spectra are less sensitive to halo parameters and
more sensitive to the dark energy.}
\label{fig:param}
\end{figure}

In the non-linear regime, the bias becomes strongly scale dependent reflecting
the correlations within a parent halo.   On the other hand, the 
ratio $b/R$ remains remarkably constant \cite{Sel00}.  Formally the combination
indicates that the galaxy-mass correlation $R>1$.  
Under the halo
model, every galaxy has a dark matter halo around it and so the
relevant power spectrum for computing the correlation coefficient is
corrected for the excess shot noise power $P^{\dgg} + \bar n_{V}^{-1}$.     
The scale dependence of $b$ and $b/R$ can be used
to pin down the halo parameters.  For fixed halo parameters, it
marks the non-linear scale which also depends on the dark energy through
the linear growth rate.

These properties remain qualitatively true for angular power
spectra with the caveat that projection effects can broaden the
non-linear transition regime for wide redshift shells for the lens galaxies
and the broad efficiency for lensing.  
In Fig.~\ref{fig:ktol} we show the correspondence
of $k$ and $l$ in the fiducial cosmology and under the Limber approximation.
In Fig.~\ref{fig:cl}, we show an example of the Limber projection to
the angular power spectra.   Note that the inflection caused by the
transition between the one host halo and (linear) two host halo regimes
occurs at wavenumbers of $l \sim {\rm few}\times 10^2$.   Note that there are
$N_L=2$ galaxy spectra, $N_L N_S=4$ cross spectra, and $N_S(N_S+1)/2=3$ shear
spectra.

In Fig.~\ref{fig:gamt}, we highlight the one halo regime by showing
the predictions for the tangential shear from Eqn.~(\ref{eqn:tangential}) 
for a given source and lens configuration.   Note that the shear
turns over as the angular radius resolves the scale radius of the halo
and shows another break on large scales marking the beginning of
the two halo regime.

\subsection{Parameter Sensitivity}
\label{sec:sensitivity}

The sensitivity of the power spectra to halo and cosmological parameters is quantified
by their derivatives with respect to the parameters in Eqn.~(\ref{eqn:derivs}).
In Fig.~\ref{fig:param} we compare the halo parameter sensitivity of galaxy-galaxy and galaxy-mass
spectra to the equation of state parameter $w_{\rm DE}$ at $z=0$ for a population with an 
abundance in the fiducial model corresponding to
$M_{\rm th}=10^{13.5} h^{-1} M_{\odot}$.   Since the initial normalization of
the power spectrum is here fixed, sensitivity to $w_{\rm DE}$ is equivalent to sensitivity to the present-day
normalization $\sigma_{8}$ (see Eqn.~(\ref{eqn:growthfunction})).  

The galaxy-galaxy spectrum is strikingly insensitive to $w_{\rm DE}$ compared with the halo parameters
that control $N_g(M_{h})$: $A_{s}$, $m_{s}$ and $\sigma_{\ln A}$.  Changes
in the parameters that 
control its shape change the power spectrum 
on all scales since they enter into the calculation of the mass threshold $M_{\rm th}$ and
hence the bias of the host halos.  A change in the cosmology
that for example lowers the present amplitude of the matter power spectrum makes galaxies of a given
number density rarer.   Since rarer objects are more highly biased tracers of the matter, 
the galaxy-galaxy power spectrum remains largely unchanged.  The same is not true
for the galaxy-mass power spectra.  Indeed the galaxy-mass power spectra are
substantially more sensitive to cosmological parameters than halo parameters.   
The combination of galaxy-galaxy and galaxy-mass power spectra is then particularly
powerful for simultaneously determining the halo and cosmological parameters.

Neither set of spectra are very sensitive to the concentration parameters
$c_s$ and $c_g$ except at very $k$ values that correspond to $l \gg 10^3$ for
the redshifts in question (see
Fig.~\ref{fig:ktol}).  By definition, 
the concentration parameters only affect the power spectrum on small scales.   
Furthermore since we have chosen a high mass threshold, the fraction of
objects that are satellites and hence affected by these parameters is low.  
This insensitivity helps justify our crude treatment of the profiles.

\begin{table}[tb]
\begin{center}
\begin{tabular}{clc}
Variable & Definition & Eq. \\
$N_L$ & Lens galaxy bins &  (\ref{eqn:Fishercross}) \\
$N_S$ & Source galaxy bins & (\ref{eqn:Fishercross}) \\
$N_h$ & Host halo occupation& (\ref{eqn:hostsatellitesplit}) \\
$N_s$ & Satellite halo occupation& (\ref{eqn:satellitenumber}) \\
$N_g$ & Total halo occupation& (\ref{eqn:hostsatellitesplit}) \\
$M$ & Halo mass& (\ref{eqn:massfn}) \\
$M_h$ & Host halo mass& (\ref{eqn:satmass})\\
$M_s$ & Satellite halo mass& (\ref{eqn:satmass}) \\
$M_v$ & Virial mass& (\ref{eqn:virialradius}) \\
$\bn$  & Angular position& (\ref{eqn:fouriertransform}) \\
$n$ & Initial tilt& (\ref{eqn:initialconditions}) \\
$n_A$ & Galaxy angular density& (\ref{eqn:na}) \\
$n_V$ & Galaxy space density & (\ref{eqn:wg}) \\
$n_h$ & Host halo space density & (\ref{eqn:massfn}) \\
$n_s$ & Satellite halo space density& (\ref{eqn:satellitemass}) \\
$w$ & Dark energy equation of state& (\ref{eqn:growth}) \\
$w_{\rm DE}$ & Constant $w(a)$&(\ref{eqn:growthfunction})  \\
$w_n$ & Specific epoch $w(a_n)$&(\ref{eqn:wn}) \\
$w_{\rm pivot}$ & Best constrained $w(a_{\rm pivot})$&(\ref{eqn:wpivot}) \\
$w_0$ & Present day $w(a=1)$&(\ref{eqn:dedom}) \\
$w_a$ & Evolution $-dw/da$&(\ref{eqn:wn})  \\
\end{tabular}
\end{center}
\caption{\footnotesize Easily confused variables.}
\label{tab:variables}
\end{table}

As for the mass-mass power spectrum, and hence the shear-shear power
spectrum, it is of course the most sensitive 
to the dark energy parameters being directly related to the linear growth function
of Eqn.~(\ref{eqn:growth}). 
It formally also depends on the satellite
mass function and hence the halo model parameters that control it.   However
we here take the perspective that the form of the mass-mass power spectrum
as a function of cosmology will in the future be fixed directly
by simulations replacing the halo model description.
Hence we take the uncertainties in the halo parameters to only affect
the galaxy-galaxy and galaxy-mass power spectra.  Operationally
we assume that variations in $P^{\dmm}$ due to the variations
in the satellite mass function are localized to the galaxy mass regime and
compensated by variations at other masses to leave the spectrum invariant.   
This corresponds to
dropping the parameter derivative terms in the Fisher matrix (\ref{eqn:Fisher}).

We provide a guide to easily confused variables
in Tab.~\ref{tab:variables}.

{}

\end{document}